\documentstyle[sprocl,psfig]{article}

\def\psizs{\psi^2(0)}
\newcommand{\be}{\begin{equation}}
\newcommand{\ee}{\end{equation}}
\newcommand{\ba}{\begin{eqnarray}}
\newcommand{\ea}{\end{eqnarray}}
\newcommand{\fr}[2]{\frac{#1}{#2}}

\newcommand{\k}{{\bf k}}
\newcommand{\q}{{\bf q}}
\newcommand{\vp}{{\bf p}}
\def\dd{{\rm d}}
\newcommand{\lb}{\left (}
\newcommand{\rb}{\right )}
\newcommand{\la}{\left\langle}
\newcommand{\ra}{\right\rangle}
\newcommand{\ep}{\epsilon}

\def\p{{\bf 1}^+}
\def\pp{{\bf 2}^+}
\def\ppp{{\bf 3}^+}
\def\pppp{{\bf 4}^+}
\def\ppppp{{\bf 5}^+}

\def\m{{\bf 1}^-}
\def\mm{{\bf 2}^-}
\def\mmm{{\bf 3}^-}
\def\mmmm{{\bf 4}^-}
\def\mmmmm{{\bf 5}^-}

\begin{document}

\title{Recent results with heavy quarks and light lepton bound
states\footnote{hep-ph/9904479, BNL-HET-99/8, TTP 99-21. 
Talk given at the 4th
Intl.~Symposium On Radiative Corrections, September 1998,
Barcelona, Spain.}}

\author{Andrzej Czarnecki}
\address{Physics Department, Brookhaven National Laboratory,\\
Upton, NY 11973}
\author{Kirill Melnikov}
\address{Institut f\"{u}r Theoretische Teilchenphysik,\\
Universit\"{a}t Karlsruhe,
D--76128 Karlsruhe, Germany}
\author{Alexander Yelkhovsky}
\address{ Budker Institute for Nuclear Physics\\
Novosibirsk, 630090, Russia}
\maketitle

\abstracts{Recent results on second order quantum corrections to
semileptonic decays of $b$ quarks and positronium spectroscopy are
reviewed.  Similarities between calculations in these seemingly
different problems are demonstrated.  Technical aspects of solving
systems of recurrence relations, and application of dimensional
regularization to bound states are explained.  }

\section{Introduction}
\label{intro}

Precision measurements play a significant role in contemporary
particle physics, spanning the energy scales starting with the
$Z$-boson pole at the high end, down to scales of the order of the $b$
and $c$ quark masses, and further down to hydrogen, positronium, and
muonium spectroscopy, and anomalous magnetic moments of leptons.  The 
role of the precision measurements reflects the successes
achieved by the Standard Model (SM) of electroweak and strong
interactions.  It has become  clear in recent years that finding any
signal of ``New Physics,'' beyond the SM expectations, is a very
difficult task.

Searching for hints about a more fundamental theory, which would
replace the SM, but lacking devices necessary to make the next step up
in energy, one resorts to a more precise analysis of the already
available data and tries to increase the accuracy of the experiments
at achievable energies.  As a consequence, more and more accurate
theoretical predictions are required practically at every point of the
energy scale up to $\sim 100$ GeV.

Theoretical studies are often impeded by complications inherent to
bound states or QCD uncertainties.  Consider for example the inclusive
decay width of the $B$-meson, $\Gamma(B \to X_c e\nu)$, used for the
$|V_{cb}|$ determination.  To calculate the decay rate in a
model-independent way, one first has to clarify how to treat the
non-perturbative bound state effects.  This non-trivial issue is
solved by invoking the Heavy Quark Expansion which permits a
calculation of the inclusive semileptonic decay width as a systematic
expansion~\cite{Bigi97} in powers of $\Lambda_{\rm QCD}/m$.  The lowest
order of this expansion corresponds to the free quark decay.  There
remains a technical difficulty of performing the standard perturbative
calculations.  Until recently, only one-loop accuracy was achievable
for charged particle decays.  Our approach to performing two-loop
calculations is one of the subjects of this talk. 

A complicated situation occurs also at the very low energies, where
Coulombic bound states like hydrogen, muonium, or positronium are
studied.  The measurements in such systems are very precise.  The
theoretical predictions, however, are complicated by the essentially
non-perturbative nature of bound state calculations.  Fortunately, it
turns out that the properties of the non-relativistic bound states,
such as energy intervals and lifetimes, can be described in terms of
series in powers and {\em logarithms} of the fine structure constant.
Here a proper organization of the calculations is a challenging
problem.

It is common in theoretical physics that problems can be solved if a
small parameter is available.  In case of the ``standard''
perturbative calculations the small parameter is usually the coupling
constant.  For the bound states, the smallness of a coupling
constant is insufficient, because the very existence of the system can
only be described in terms of an infinite number of Feynman diagrams.
The situation is simplified in case of non-relativistic bound states,
which encompass the light atoms and ``leptonia'' for which precision
measurements are  being performed.  There, the relative velocity of the
bound state components is small and can serve as an expansion
parameter.  

For the bound states, expanding around a non-relativistic limit is the
only way known at present to arrive at precise results.  But even in
case of the more conventional perturbative calculations, Feynman
integrals often are intractable and the only option to
evaluate them is to {\em invent} a small parameter.  It is often
convenient  to use some kinematic parameter, so that gauge
invariance and other symmetries of the problem are preserved.

Construction of non-conventional expansions of Feynman graphs has
recently helped to solve many outstanding problems in high precision
theoretical studies.  In what follows we are going to describe some
examples of how such expansions can be constructed, and review our
recent calculations devoted to the semileptonic decay width of the $b$
quark and the positronium energy levels.  We deliberately focus on the
technical aspects of those calculations, to provide additional
information and supplement our journal publications.

\section{Inclusive semileptonic decay width of the $b$ quark}

\begin{figure} 
\hspace*{-19mm}
\begin{minipage}{16.cm}
\vspace*{3mm}
\[
\mbox{
\psfig{figure=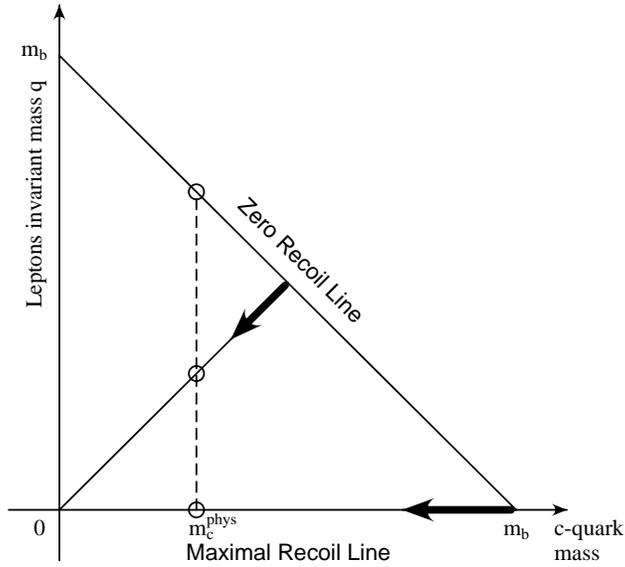,width=80mm,bbllx=25pt,bblly=104pt,%
bburx=582pt,bbury=630pt} 
}
\]
\end{minipage}
\caption{Status of the two-loop QCD corrections to the decay $b\to
c+\mbox{leptons}$.  The dashed line denotes the physical region for
the actual $c$ quark.  Points where the full corrections are known are
circled.  An analytical formula is known along the whole zero recoil
line.~\protect\cite{zerorecoilA,Franzkowski:1997vg}  The other two
points are found from expansions around the base points of the two
arrows: at maximal recoil~\protect\cite{Czarnecki:1997hc} and at the
intersection with the diagonal.~\protect\cite{b2cHalf}}
\label{fig:triangle}
\end{figure}

Let us start by discussing a calculation of the second order QCD
corrections, ${\cal O}(\alpha_s^2)$, to the semileptonic decay width
of the $b$ quark, $\Gamma(b \to c e \nu_e)$. At present a complete
calculation is not feasible.  However, one can calculate corrections
at this order to the differential decay width ${\rm d}\Gamma/{\rm
d}q_{e \nu}^2$ for special values of the leptonic invariant mass,
$q_{e \nu}^2 \equiv q^2$.  Such calculations were performed at three
points:
\begin{itemize}
\item
$q^2=(m_b-m_c)^2$ (zero recoil);~\cite{zerorecoil,zerorecoilA}
\item
$q^2=m_c^2$ 
(intermediate recoil);~\cite{b2cHalf}
\item
 $q^2=0$ (maximal recoil).~\cite{Czarnecki:1997hc}
\end{itemize}
  These kinematical configurations are depicted in 
Fig.~\ref{fig:triangle}.
Based on these 3 results, we performed a fit to interpolate the
corrections for arbitrary $q^2$.  From that fit, a correction to the
inclusive decay rate, $\Gamma(b \to c e\nu_e)$, was estimated.

For kinematical configurations away from the zero recoil limit one
should consider three different types of corrections: pure virtual
corrections, single gluon radiation, and radiation of two gluons (at
zero recoil only the first type is present).  Our approach is to
calculate these contributions separately and then sum them up in the
final result.\footnote{An alternative approach to such calculations is
based on the optical theorem; to get an $n$-loop result, one
calculates the imaginary part of a certain $(n+1)$-loop Feynman
integral. Such method was applied e.g.~to calculate QED corrections to
the muon decay rate.~\cite{vanRitbergen:1998yd}} Below we describe the
essential ingredients of those calculations.

\subsection{Virtual corrections}
\label{virtual}
As an example of the virtual correction calculation we consider the
intermediate recoil case, equating the invariant mass of the lepton
system, $\sqrt{q^2}$, to the $c$-quark mass.  If the $c$ quark had
exactly half the $b$ quark mass, this configuration would coincide
with the zero-recoil limit (see Fig.~\ref{fig:triangle}), in which
case the virtual corrections are known
exactly.~\cite{zerorecoilA,Franzkowski:1997vg} For the actual $c$ and
$b$ masses these corrections can be calculated by expanding the
amplitude in power series in $(m_b-2m_c)/m_b$, that is around the
known zero recoil case.  An important aspect of this expansion is that
it commutes with the virtual momenta integrations; it simply amounts
to Taylor expanding the integrand.

After the expansion, all integrals we need are of the
zero-recoil type.  They can be divided up into the following categories
(they are written here in a general form, with $\omega$ denoting the
ratio of the final and the initial quark masses.  
For our intermediate recoil calculation we need only $\omega=1/2$):  
\begin{enumerate}
\item Planar with five propagators
\ba
\lefteqn{P_5(a_1,a_2,a_3,a_4,a_5) = \pi^D }
\nonumber \\
&&\times\int 
{ {\rm d}^D k_1 {\rm d}^D k_2 
\over
 k_1^{2a_1} (k_1-k_2)^{2a_2} k_2^{2a_3} 
 \left( k_1^2 + 2k_1p\right)^{a_4}
 \left( k_2^2 + 2k_2p\omega \right)^{a_5}
}.
\label{eq:defPlanar}
\ea
\item Non-planar, first type, five propagators
\ba
\lefteqn{N_P(a_1,a_2,a_3,a_4,a_5) = \pi^D 
\int 
{ {\rm d}^D k_1 {\rm d}^D k_2 
\over
 k_1^{2a_1}  k_2^{2a_2} }}
\nonumber \\
&& \times \frac {1}{
 \left( k_1^2 + 2k_1p\right)^{a_3}
 \left( k_2^2 + 2k_2p\omega \right)^{a_4}
 \left[ (k_1+ k_2)^2 + 2(k_1+k_2)p \right]^{a_5}
}
\label{eq:defNP}
\ea

\item Non-planar, second type, five propagators

\ba
\lefteqn{N_P'(a_1,a_2,a_3,a_4,a_5) = \pi^D \int 
{ {\rm d}^D k_1 {\rm d}^D k_2 
\over
 k_1^{2a_1}  k_2^{2a_2} }
}
\nonumber \\
&&\times \frac {1}{
 \left( k_1^2 + 2k_1p\right)^{a_3}
 \left( k_2^2 + 2k_2p\omega \right)^{a_4}
 \left[ (k_1+ k_2)^2 + 2\omega(k_1+k_2)p \right]^{a_5}
}
\label{eq:defNP1}
\ea

\item Non-planar with six propagators
\ba
\lefteqn{N_6(a_1,a_2,a_3,a_4,a_5,a_6) = \pi^D 
\int 
{{\rm d}^D k_1 {\rm d}^D k_2 
\over
 k_1^{2a_1}  k_2^{2a_2} 
 \left( k_1^2 + 2k_1p\right)^{a_3}}
}
\nonumber 
\\
&&\hspace*{-10mm}
\times \frac {1}{
 \left[ (k_1+ k_2)^2 + 2(k_1+k_2)p \right]^{a_4}
\left[ (k_1+ k_2)^2 +2\omega(k_1+k_2)p \right]^{a_5}
 \left( k_2^2 +2\omega k_2p \right)^{a_6}
}
\nonumber \\
\label{eq:defN6}
\ea
\end{enumerate}

Taylor expansion of the amplitude may result in very many terms,
differing by powers of the propagators, $a_i$.  It is crucial to find
a general algorithm of calculating them, so that an algebraic
manipulation program can be employed.  It is convenient to use
recurrence relations derived using integration-by-parts,~\cite{che81} to
reduce an arbitrary integral to a small basis set of the so-called
master integrals. 

Let us note at this point that solving recurrence relations is still
more of an ``art'' than ``science.''  It would be very desirable to
have a better understanding of the mathematical structure of the
systems of recurrence relations; a very promising approach to this
problem is being developed by P. Baikov.~\cite{Baikov:1996rk}

Here, we would like to demonstrate our approach to solving such
recurrence relations with the example of the planar integral
$P_5(\{a_i\})$.  First, a system of relations is derived in the usual
way~\cite{che81} by equating integrals of total derivatives to zero.
One finds
\ba
M_1 &=& D-2a_1-a_2-a_4+a_2\pp(\mmm-\m)-a_4\pppp\m
\nonumber\\[2mm]
M_2 &=&
D-a_1-2a_2-a_4+a_1\p(\mmm-\mm)
+a_4\pppp\left({\omega-1\over\omega}\mmm-\mm
 +{1\over \omega}\mmmmm\right)
\nonumber\\[2mm]
M_3 &=& D-a_1-a_2-2a_4-a_1\p\mmmm
+a_2\pp\left({\omega-1\over\omega} \mmm-\mmmm+{\mmmmm\over\omega}\right)
+2a_4\pppp
\nonumber\\[2mm]
M_4 &=& D-2a_2-a_3-a_5+a_3\ppp(\m-\mm)
+a_5\ppppp\left(
 (1-\omega)\m-\mm+\omega\mmmm\right)
\nonumber\\[2mm]
M_5 &=&
2D-2a_1-2a_2 -2a_3 -a_4-a_5 -a_4\pppp\m-a_5\ppppp\mmm
\nonumber\\[2mm]
M_6 &=&
D-a_2-a_3-2a_5+a_2\pp\left((1-\omega)\m+\omega\mmmm-\mmmmm\right)
\nonumber\\  && \qquad \qquad 
-a_3\ppp\mmmmm +2\omega^2a_5\ppppp
\nonumber\\[2mm]
M_7 &=&
\left({3D\over 2}-\sum_{i=1}^5 a_i\right)(\m-\mmmm)
+2a_4\pppp\m+a_5\ppppp \omega(\m-\mm+\mmm)
\nonumber\\[2mm]
M_8 &=&
{1\over \omega}
\left({3D\over 2}-\sum_{i=1}^5 a_i\right)(\mmm-\mmmmm)
+a_4\pppp(\m-\mm+\mmm) +2\omega a_5\ppppp \mmm
\label{eq:system}
\ea
The recursive algorithm proceeds in three steps: 
\begin{enumerate}
\item
the exponents of the massive propagators ($a_{4,5}$) are reduced to 1; 
\item
the exponent of the massless propagator containing both integration
impulses ($a_2$) is reduced to 1;  
\item 
the exponents $a_{1,3}$ are reduced.
\end{enumerate}
The first step is simple.  We assume that $a_4$ and $a_5$ are
positive; otherwise the integral can be done by a sequence of one-loop
integrations.  We use the relation $M_6$.
The term $2\omega^2 a_5\ppppp$ involves an integral with $a_5$ higher
than any other term in this relation.  Therefore we can use $M_6$ 
to express $P_5(\ldots,a_5+1)$ by $P_5(\ldots,a_5)$.
An analogous manipulation is done using $M_3$ to reduce $a_4$. 
Eventually we obtain $a_4,a_5 \le 1$. 

The next step is accomplished as follows. 
We first would like to eliminate all operators that contain $\pppp$ and
$\ppppp$ from the system (\ref{eq:system}).  There are eight such
operators $\pppp$, $\pppp\m$, $\pppp\mm$, $\pppp\mmm$, and $\ppppp$,
$\ppppp\m$, $\ppppp\mm$, $\ppppp\mmm$.  An important point is that
this number is equal to the number of the recurrence relations and we
can therefore solve the recurrence relations for these operators.

We denote the resulting 8 equations by $T_{4,41,42,43,5,51,52,53}$.
The notations are such that for example $T_{41}$ means an equation
which expresses the operator $\pppp\m$ through other operators.  These
equations allow us to obtain new recurrence relations using 
$a_4( T_{41} - \m T_4) = 0$ and similar equations.  In
particular, we use a relation $a_4( T_{43} - \mmm T_4)=0$ to
lower the value of $a_2$ to $1$.

After this has been achieved, we eliminate all operators which contain
$\pp$.  At this stage, the problem is essentially two parametric
(i.e.~we have to move in the $a_1-a_3$ plane).  Finally, we are able
to reduce any $P_5(\{a_i\})$ integral to four integrals
$P_5(1,1,1,1,1)$, $P_5(0,1,1,1,1)$, $P_5(-1,1,0,1,1)$, and
$P_5(0,1,0,1,1)$ and a number of trivial integrals.  The four master
integrals are
\ba
P_5(1,1,1,1,1)&=&-\frac {\pi^2}{3\ep} -\frac {2 \pi^2}{3} 
- \frac{5\zeta (3)}{2} - \pi^2\ln 2
\nonumber \\
P_5(0,1,1,1,1) &=& \frac {1}{2\ep^2} + \frac {1}{\ep}\left( \frac {5}{2} 
+  2\ln2 \right ) + \frac {19}{2} + 10\ln 2
                    - \frac {5}{12}\pi^2
\nonumber \\
P_5(-1,1,0,1,1) &=& 
        \frac {19}{16\ep^2} 
       + \frac {1}{\ep} \left ( \frac {311}{96} + \frac {3\ln 2}{4}
        \right )
+ \frac {1225}{192} + \frac {79}{24}\ln 2 - \frac {\ln^2 2 }{2} 
            + \frac {15 \pi^2}{32}
\nonumber \\
P_5(0,1,0,1,1) &=& 
       - \frac {5}{8\ep^2} 
       - \frac {1}{\ep} \left (  \frac {29}{16} + \frac {\ln 2}{2}  \right)
       - \frac {107}{32} - \frac {13}{4}\ln 2 + \ln^2 2 
- \frac {11\pi^2 }{48}.
\nonumber
\ea
The above approach to solving recurrence
relations is quite general.  One sees that such solution cannot be
constructed in a 
completely universal way.  However, our approach allows to go through
a number of steps, constructing a solution without much effort and it
has proved useful in various cases.  In particular, it works for the
``hard'' threshold integrals, necessary for the calculations described
in Section \ref{sec:hard}.

\subsection {Emission of two gluons in semileptonic $b$ decays}

For simplicity, we will discuss here the case of the maximal recoil,
$q^2=0$. The intermediate recoil case is treated in an analogous way.
A more detailed discussion of this issue is given in
\cite{maxtech}. Here the most essential features are summarized.

The basic observation in this part of the calculation is that the
energy losses caused by radiation off a non-relativistic particle are
small. This is evident in the Coulomb gauge, since the interaction of
magnetic photons with charged particles is proportional to the
velocity of the latter. Therefore, the physical properties of the real
radiation provide a hint how an expansion should be organized.
Considering a decay $b \to c +e + \nu_e +g_1 + g_2$ in the case
$q^2\equiv (e+\nu_e)^2 =0$, we intend to calculate it expanding in
$\delta = (m_b-m_c)/m_b$.  For brevity we will refer to the $e,\nu$
system as to the $W$ boson.  Taking into account phase space
constraints, it is easy to establish that the spatial momentum of the
$c$ quark, as well as the four-momenta of the $W$, $g_1$ and $g_2$ are
of the order $\delta$.  Then, a natural expansion of the propagators
emerges: all propagators of the virtual particles can be expanded
around the {\em static limit}, which essentially simplifies the
calculation. Clearly, such an expansion will generate a large number
of terms, but this can be handled using symbolic manipulation
programs.

The last non-trivial question is the integration of the obtained
expression over the four particle phase space. Given that the
remaining propagators are static and the $W$ boson does not
participate in the interaction, it is clear that the integration over
two-particle phase (sub)space ($W,c$) can be easily performed.  The
remaining part of the calculation requires an integration over the
three particle phase space -- two gluons and the ``particle'' composed
of $W$ and $c$.  A systematic expansion of this object is accomplished
by an appropriate choice of variables, described in detail in
\cite{maxtech}.

\subsection{One gluon radiation and one virtual correction}

Single gluon emission in the presence of a virtual loop is, in
principle, possible to compute exactly. However, it is very inconvenient,
e.g.~because of necessity of integrating four-point functions over a
three particle phase space.

We have adopted a different approach, based again on an expansion of
such amplitudes around the static limit ($m_b=m_c$).  In contrast to
the virtual corrections and double gluon emission, in the present case
the expansion is non-trivial.  Let us consider the following diagram:
first a virtual gluon is exchanged between $c$ and $b$ quarks, and
then the $c$ quark emits a real gluon.  An essential point is that
before the real gluon emission the $c$ quark is {\em off mass shell},
which provides an infra-red regularization for the virtual loop.
Expanding the one-loop amplitude naively, i.e.~putting the
``intermediate'' $c$-quark line loop on shell, we change the infrared
properties of the diagram and divergences will result in singularities
in $1/(D-4)$ (we use dimensional regularization).  To avoid these
spurious singularities the expansion should be ``corrected''; in this
case the so-called eikonal expansion~\cite{CzarSmir96,Smirnov96} is
used.  In the context of the two loop corrections to fermion decays
this expansion is described in detail in Ref.~\cite{maxtech}.

\subsection{ General remarks on semileptonic heavy fermion decays}

Let us summarize the technical experience we have gained by studying
semileptonic $b$ and $t$ decays to second order in the strong coupling
constant.

First, we note that the problem was made tractable by inventing a ``small''
parameter.  For the top quark decays~\cite{Czarnecki:1998qc} the 
result is obtained if we take the ``small'' parameter close
to unity.  This clearly pushes the applicability of the method to its
limits.  However, even in this extreme case reasonable numerical
estimates can be obtained.

Second, because we consider processes with the on--shell external
particles, one has to be careful in analyzing contributing regions of
integration momenta.  In particular, one can encounter non-analytical
dependence on the expansion parameter. In many cases the relevant
momentum regions can be found by considering the corresponding
effective theories.

Third, it is possible to construct an expansion of the amplitudes with
real radiation of massless particles starting from the
non-relativistic limit.  It is likely that this construction can be
applied beyond the semileptonic decay width, for example in a
perturbative calculation of the cross section of such process as
$e^+e^- \to \mu^+ \mu^- \gamma \gamma$ close to the two muon threshold,
or even for higher energies if one takes into account sufficiently
many terms of such an expansion.

\section{Bound States in QED}

Consistent description of bound states is one of the most complicated
problems in Quantum Field Theory.  The problem is simplified if the
bound state components are non-relativistic.  This is the case, for
example, in QCD in $\Upsilon$ or $\psi$ meson families, or in QED in
light atoms and in leptonic bound states like positronium or muonium.

The difficulty in the study of bound states is their essentially
non--per\-tur\-bative nature, as we discussed in Section \ref{intro}.
In the non-relativistic case the situation is simplified: to first
approximation one can describe the bound state using a Schr\"odinger
wave function.  This description can subsequently be refined by means
of perturbation theory, using the relative velocity $v$ of the
components as an expansion parameter. 

This kind of expansions can be constructed both for QED and QCD.  In
what follows, we limit ourselves to the QED where a low-energy
effective theory approach is well established, and is known as
Non-Relativistic Quantum Electrodynamics (NRQED).~\cite{Caswell:1986ui}

\subsection{NRQED}

Very roughly speaking, the idea underlying NRQED consists in
separating the soft scale $\sim m\alpha$, which characterizes the
bound state, from the hard one $\sim m$, and ``integrating out'' the
latter.  The interactions at the soft scale can be described by
Quantum Mechanics, i.e.~employing a non-relativistic approximation.
The physics at hard scales is governed by the relativistic QFT.
However, when we are interested in processes which occur at small
momenta and energies, the contribution of hard scales can be accounted
for by introducing effective local operators and computing their
coefficients in the effective Hamiltonian.

In the lowest order of this construction we describe the system by a
bound state wave function.  For positronium it is found from
the Schr\"odinger equation with the ``leading order'' Hamiltonian ($m$
is the electron mass, $r$ is the distance between the electron and the
positron):
\be
H_{\rm LO} = \frac {\vp ^2}{m}-\frac {\alpha}{r}.
\label{eq:loh}
\ee
In a 3-dimensional space the ground state wave function (in momentum
representation) is 
\be
\phi(p) = \sqrt{\pi\alpha m\over 2}{2m^2\alpha^2\over (\vp^2-mE)^2},
\qquad
E=-{m\alpha^2\over 4}.
\label{wf}
\ee
The difference between the full Hamiltonian and Eq.~(\ref{eq:loh}) can
be treated as a perturbation.  The difficulties of calculating in
NRQED are caused by divergences in the matrix elements of the
subleading operators. In the original formulation~\cite{Caswell:1986ui}
the ultraviolet divergences were dealt with by subtracting
counterterms defined with help of diagrams evaluated on-shell and at
threshold.  Those diagrams, in turn, contain infrared divergences,
which are regularized by introducing a small mass $\lambda$ for the
photon.

Since that pioneering work it has become clear that working with a
massive photon leads to technical difficulties, particularly in
evaluating the so-called hard corrections, arising at scales $\sim m$.
Indeed, the hard corrections remain unknown (at two-loops) for such
processes as para or orthopositronium decays, while the soft scale
effects have already been evaluated.  The reason for this is that the
two-loop diagrams are difficult to evaluate when more than one mass
scale is present.  Although the only relevant scale in hard
corrections for positronium physics is the electron mass,
regularization with a photon mass introduces a second scale $\lambda$.
The hard corrections can, however, be computed in many configurations,
if instead of introducing $\lambda$ one uses dimensional
regularization.

It has been shown~\cite{Pineda:1997bj} that dimensional regularization
permits an exact separation of effects arising at various
characteristic energy scales.  Using that method, the complete energy
spectrum of Ps has been reproduced~\cite{Pineda:1998kn} to order
$m\alpha^5$.  Further, we have computed analytically $m\alpha^6$
corrections to the hyperfine splitting (HFS) of the Ps ground
state.~\cite{Czarnecki:1998zv}  More recently,~\cite{levels} we
generalized that result to all $S$ states and computed also their spin
independent shift at ${\cal O}(m\alpha^6)$.  Here we would like to
discuss some features of the NRQED in dimensional regularization,
which we found useful in those calculations.

\subsection{Quantum Mechanics in $d$ dimensions}

The first point we would like to discuss is a construction
of the lowest order approximation. We
consider the Schr\"odinger
equation in a $d$-dimensional  momentum space,
\be
\phi(p) = \fr {4\pi \alpha m}{\vp^2-mE}
\int \fr {{\rm d}^{\rm d} \k}{(2\pi)^d} 
\fr {\phi(k)}{(\vp-\k)^2}.
\label{ddS}
\ee
The solution of this equation for arbitrary $d$ is not known.
This complicates the calculations which involve the wave function.
However, since in the configuration space all divergences
are located at $r=0$, it is possible to extract divergent 
contributions in the form of $\psi^2(0)/\ep$.  Thus, the unknown
overall factor
$\psi^2(0)$ will be present in the intermediate results.  In the final
result the singularities $1/\ep$ cancel and we can replace $\psi(0)$
by its 3-dimensional value.  

In rare cases we need, in addition to the value of the configuration
space wave function at  the origin, also the behavior of its first
derivative.  This, however, is related to $\psi(0)$ by the
Schr\"odinger equation.  To demonstrate this, we consider
the Schr\"odinger equation in the configuration space.  Since for $S$
states $\psi(r) $ has no angular dependence, we have
\ba
\left( -{\nabla^2\over m} + V(r)\right) \psi(r) 
 = -{1\over m}\left( 
 \partial_r^2 + {d-1\over r} \partial_r - m V(r)\right) \psi(r)
= 0,
\label{eq:ddS}
\ea
where  $V(r)$ is the Coulomb potential in $d$ dimensions,
\be
V(r) = - {\alpha \Gamma\left({d\over 2} -1\right) \over 
             \pi^{{d\over 2}-1} r^{d-2}}.
\ee
We assume that the wave function near the origin behaves like
$
\psi(r) \simeq \psi(0) ( 1-cr^\beta),
$
and find $c$ and $\beta$ by substituting this form into
Eq.~(\ref{eq:ddS}):
\be
\psi(r\to 0) \simeq \psi(0) \left( 1 
- {m\alpha\Gamma\left({1\over 2} - \ep\right)
   \over 2\pi^{{1\over 2}-\ep}(1+2\ep)}r^{1+2\ep}
                       \right). 
\ee

We now would like to demonstrate how divergent integrals are treated
in practice, using two examples from our hyperfine splitting
calculations.  First, we consider a logarithmic divergence
arising from tree-level Coulomb photon exchange,~\cite{levels}
\be
\Delta_C E_{\rm hfs} 
= - \fr{\pi\alpha}{dm^4}
\left\langle \fr{ (\vp'\q)(\q\vp) }{ \q^2 } \right\rangle.
\label{int}
\ee
In Eq.~(\ref {int}) the  matrix element is to be calculated over the ground 
state wave function in $d$ dimensions:
$$
\la f(\vp,\vp') \ra \equiv 
\int {\dd^d p \over (2\pi)^d}{\dd^d p' \over (2\pi)^d}
\phi(p)\phi(p') f(\vp,\vp').
$$
Although the integrand does not look complicated, the
difficulty is that the exact form of the wave function $\psi(r)$ in
$d$ dimensions is not known.  We use the fact that it 
satisfies the $d$-dimensional Schr\"odinger
equation,  Eq.~(\ref{ddS}).
Using that equation we rewrite the integral in Eq.~(\ref {int}) as
\be
\la \fr{ (\vp'\q)(\q\vp) }{ \q^2 } \ra _{\vp,\vp'} =  
\la \fr{(4\pi\alpha m)^2(\vp'\q)(\q\vp) }
{(\vp^2-mE)(\vp-\k)^2 \q^2 (\vp'^2-mE)(\vp'-\k')^2}
\ra _{\k,\k'},
\label {int1}
\ee
where the integration over $\vp$, $\vp'$, as well as $\k$, $\k'$, in the
last expression is 
understood.  The integral over $\vp$ and $\vp'$ receives a divergent
contribution only from the region where $\vp$ and $\vp'$ become
simultaneously infinite. Therefore, a single subtraction is sufficient
to make this integral finite.  It is convenient to subtract from
(\ref{int1}) the following expression: 
\be
\la \fr{(4\pi \alpha m)^2  (\vp'\q)(\q\vp) }{(\vp^2-mE)^2 \q^2
(\vp'^2-mE)^2}  \ra _{\k,\k'}.
\label {int2}
\ee
After the subtraction is done, two nice features emerge.  In Eq.~(\ref
{int2}) the integration over $\k,\k'$ factorizes and leads to
$\psizs$ times a two-loop integral, which can be easily calculated
for arbitrary $d$.  On the other hand, the difference between the last
integral in Eq.~(\ref {int1}) and the integral in Eq.~(\ref {int2}) is
finite and can be calculated for $d=3$ using the explicit form of the
wave function, Eq.~(\ref{wf}).

We note that the counterterm (\ref {int2}) is
constructed in such a way that the difference mentioned above vanishes
for the ground state. This can be easily seen by integrating over
$\k,\k'$ in Eq.~(\ref {int2}) and noticing that the
$\vp,\vp'$-dependent terms in the denominator of Eq.~(\ref {int2})
coincide (up to a normalization factor) with the three-dimensional
ground state wave functions in the momentum representation.

Finally, for the Coulomb exchange contribution one finds
\be
\Delta_C E_{\rm hfs} = \fr{ \pi\alpha^3 }{ 24 m^2 } \psizs
                \lb \fr{ 1 }{\epsilon } - 4 \ln(m\alpha) - \fr{1}{3} \rb,
\label{DeltaC}
\ee

Our second example is a linearly divergent integral arising from the
tree-level magnetic photon exchange,~\cite{levels}
\ba
\la \vp^2 \ra &=& \psi(0) 
\int \fr{\dd^d p}{(2\pi)^d} \vp^2 \phi(p)
\nonumber \\
&=& m \psi(0)  
\int \fr{\dd^d p}{(2\pi)^d} \lb E \phi(p) 
+  
\int \fr{\dd^d k}{(2\pi)^d} \fr{ 4\pi\alpha }{ (\vp-\k)^2 } 
\phi(k) \rb.
\label {tadpole}
\ea
Shifting the integration variable $\vp \to \vp+\k$ we find that the
$\vp$-integral in the last term is scale-less.  In dimensional
regularization such integrals vanish.  The first term in Eq.~(\ref
{tadpole}) is finite in three dimensions. We obtain
\be
\la \vp^2 \ra = mE \psizs.
\ee

\subsection{Hard scale contributions}
\label{sec:hard}
Perhaps the main advantage of using dimensional regularization in
bound state calculations is the possibility of computing
the hard corrections analytically.

The contributions of hard scales arise from virtual momenta of
order of the electron mass.  They can be calculated by considering
the on--shell $e^+e^-$ scattering amplitude  exactly at the
threshold, i.e.~for zero relative velocity of the incoming electron
and positron.  Since the amplitude is independent 
of the incoming spatial momenta,
it gives rise to four-fermion operators in the low-scale Lagrangian or,
equivalently, to the $\delta(\mbox{\boldmath$r$})$ terms in the
effective quantum mechanical Hamiltonian.

Technically, this calculation is similar to the derivation of 
the matching coefficient of the vector quark-antiquark current in QCD
and its NRQCD counterpart.~\cite{threshold,BSS}

An arbitrary Feynman integral which contributes to the hard scale part of
the calculation can be written as
\be
I(a_1,...a_9) = \int \fr {\dd^D k_1}{(2\pi)^D} \fr {\dd^D k_2}{(2\pi)^D}
{1\over S_1^{a_1}S_2^{a_2}S_3^{a_3}S_4^{a_4}S_5^{a_5}S_6^{a_6}S_7^{a_7}
S_8^{a_8} S_9^{a_9}},
\ee
where
\ba
&&
S_1 = {k_1^2},\qquad   
S_2={k_2^2},\qquad   
S_3={(k_1-k_2)^2}, \qquad    
S_4 = {k_1^2+2pk_1},
\nonumber \\
&&
S_5= {k_2^2+2pk_2},\qquad
S_6 = {k_1^2-2pk_1},\qquad    
S_7 = {k_2^2-2pk_2},
\nonumber \\
&&
S_8 = {(k_1-k_2)^2+2p(k_1-k_2)},\qquad
S_9 = {(k_1-k_2)^2-2p(k_1-k_2)},
\ea
and $a_1,\ldots,a_9$ are integers.  
In practice we find diagrams with only at most 6 different
propagators, so that at least 3 exponents $a_i$ are zero.

A method for computing these integrals is analogous to the 2-loop
virtual corrections to quark decays.  We have described our algorithm
in Section \ref{virtual}.  

\section{Conclusions}
In this talk we have presented technical aspects of our recent
calculations of two-loop corrections to semileptonic quark decays and
positronium properties.  The methods developed in the context of quark
decays, in particular the algorithm for solving systems of recurrence
relations, have found applications  in positronium
physics.  We hope that further progress can be made in such atomic
physics calculations using the tools developed for high energy
problems.  In particular, applying dimensional regularization should
facilitate so far intractable calculations, such as two-loop
corrections to positronium decays.


\end{document}